\documentstyle[epsf,twocolumn,prl,aps]{revtex}

\begin{document}
\draft

 \twocolumn[\hsize\textwidth\columnwidth\hsize\csname @twocolumnfalse\endcsname
\title{Competition Between Stripes and Pairing in a $t-t^\prime-J$ Model}

\author{ Steven R.\ White$^1$ and D.J.\ Scalapino$^2$}
\address{ 
$^1$Department of Physics and Astronomy,
University of California,
Irvine, CA 92697
}
\address{ 
$^2$Department of Physics,
University of California,
Santa Barbara, CA 93106
}
\date{\today}
\maketitle
\begin{abstract}
\noindent

As the number of legs $n$ of an $n$-leg, $t-J$ ladder increases, density
matrix renormalization group calculations have shown that the doped state
tends to be characterized by a static array of domain walls and that
pairing correlations are suppressed.  Here we present results for a
$t-t^\prime-J$ model in which a diagonal, single particle, next-near-neighbor
hopping $t^\prime$ is introduced. We find that this can suppress the 
formation of
stripes and, for $t^\prime$ positive,
 enhance the $d_{x^2-y^2}$-like pairing correlations. The effect
of $t^\prime > 0$ is to cause the stripes to evaporate into pairs and for
$t^\prime <0$ to evaporate into quasi-particles. Results
for $n=4$ and 6-leg ladders are discussed.

\end{abstract}


 ]



Neutron scattering experiments on 
La$_{1.6-x}$Nd$_{0.4}$ Sr$_x$CuO$_4$
show evidence of a competition between static (quasi-static) stripes and
superconductivity\cite{tra97}. 
Here the stripes consist of (1,0) domain walls of holes
separating $\pi$-phase shifted, antiferromagnetic regions. For $x=0.12
\ (x\approx 1/8)$, the intensity of the charge and spin superlattice peaks is
largest and $T_c$ is less than 5K. As $x$ deviates from this value,
the relative intensity of the magnetic superlattice peaks
decrease and the superconducting transition temperature
$T_c$ increases. High field magnetization studies \cite{ost97} provide
evidence that in this material superconducting can coexist with static (or
quasi-static) stripe order. However, the fact that $T_c$ is a minimum
where the superlattice peaks are most intense suggests
that static stripe order competes with
superconductivity. 

We are interested in understanding whether a $t-J$-like model
can exhibit this type of behavior. In studies of $n$-leg, $t-J$
ladders we have previously found evidence for stripe formation.
In particular, for $n=$ 3 and 4 legs we have found evidence for
both stripes and pairing \cite{WS97,WS98}. These systems have
open boundary conditions. However, in wider ladders ($n=6$ and
$n=8$) with cylindrical boundary conditions, where the stripes
close on themselves rather than having free ends, the stripes
appeared to be more static and the pairing correlations were
found to be suppressed\cite{WS98b}.  This suppression of the
pairing correlations was also observed when an external
potential was applied to further pin the stripes. 

If the formation of static stripes could be suppressed, one
might hope to find enhanced pairing correlations. It is not
clear whether the complete elimination of stripes or only a
slight destabilization would be more favorable to pairing
correlations.  We have been investigating various
interaction terms which could destabilize stripes. Here we focus
on the effect of a next-near-neighbor diagonal hopping
$t^\prime$.  Effective hopping parameters have
been evaluated from band structure calculations and finite CuO
cluster calculations. For the hole-doped cuprates $t^\prime$ is
found to be negative while for the electron-doped cuprates it is positive.
Both $t^\prime$ and the one-electron hopping $t^{\prime\prime}$,
which connects next-near-neighbor sites along the (0,1) or (1,0)
axis, have been used in $t-t^\prime-t^{\prime\prime}-J$ models to
fit ARPES data \cite{kim98}. In addition, Lanczos calculations
by Tohyama and Maekawa\cite{TM94} on $t-t^\prime-J$ clusters and Monte
Carlo calculations\cite{DM95} on $t-t^\prime$ Hubbard lattices show that
$t^\prime>0$ tends to stabilize the commensurate $(\pi,\pi)$
antiferromagnetic correlations. Recently, exact
diagonalization and density-matrix
renormalization group (DMRG) calculations on small clusters and
four-leg ladders have found that $t' < 0$ destabilizes 
stripes\cite{tohyama}. Furthermore, it was concluded that a small
positive $t'$ did not destabilize the stripes on these systems.

Here we will consider the effect of $t'$ on 
both open four leg and cylindrical six leg ladders. In addition
to considering the affect of $t'$ on stripe stability, we will
measure its affect on pairing correlations. We find that stripes
are destabilized for either sign of $t'$, and that pairing is
suppressed for $t'<0$, and enhanced for $t'>0$. This latter
effect is surprising, since superconducting transition
temperatures are generally higher for hole doped cuprates ($t' < 0$)
than for electron doped ($t' > 0$).

The $t-t^\prime-J$ Hamiltonian
which we will study has the form 
\begin{eqnarray}
H=-t\sum_{\langle ij\rangle s} &(c^+_{is} c_{js} + c^+_{js} c_{is}) 
- t^\prime \sum_{\langle ij\rangle' s}
(c^+_{is} c_{js} + c^+_{js}  c_{is})\nonumber \\  
&+ J\sum_{\langle ij\rangle} (\vec S_i \cdot \vec S_{j} -
\frac{1}{4}\, n_i n_j) \label{oneone}
\end{eqnarray} 
Here
$\langle ij\rangle$ are near-neighbor sites, $\langle
ij\rangle^\prime$ are diagonal next-near-neighbor sites, $\vec
S_i = \frac{1}{2} c^+_{is} \sigma_{ss^\prime} c_{is}$, $n_i =
c^+_{i\uparrow} c_{i\uparrow} + c^+_{i\downarrow}
c_{i\downarrow}$, and $c^+_{is}(c_{is})$ creates (destroys) an
electron of spin $s$ at site $i$. No double occupancy is
allowed. We will use DMRG calculations to explore the charge,
spin, and pairing correlations on doped four and six leg ladders.
The calculations reported below keep up to 1200 states per
block, with truncation errors of about $10^{-4}$, and from six to ten
finite system sweeps. We have checked the inclusion of $t'$ in 
our program by comparing the results for the rung hole density
on a $14\times4$ system with the results of Tohyama, et. al.
\cite{tohyama}; precise agreement was found.

In a previous study of the 4-leg $t-J$ ladder, we found that four-hole
diagonal domain walls formed as the doping increased. 
In Figure 1(a) and (b) we show the rung density
\begin{equation}
\left\langle n_r(\ell)\right\rangle = \sum^4_{i=1}
\, \left\langle n_{\ell i}\right\rangle
\label{onetwo}
\end{equation}
versus $\ell$ for $J/t=0.35$ on a $12\times 4$ lattice with 8
holes and open boundary conditions. For $t^\prime=0$, we clearly 
see the formation of two domain walls, signaled by two broad
peaks in $\left\langle n_r(\ell)\right\rangle$.
As $t^\prime/t$ is varied, one clearly
sees that the static domain wall structure is suppressed for
either sign of $t^\prime$.

\begin{figure}[ht]
\epsfxsize=3.3 in\centerline{\epsffile{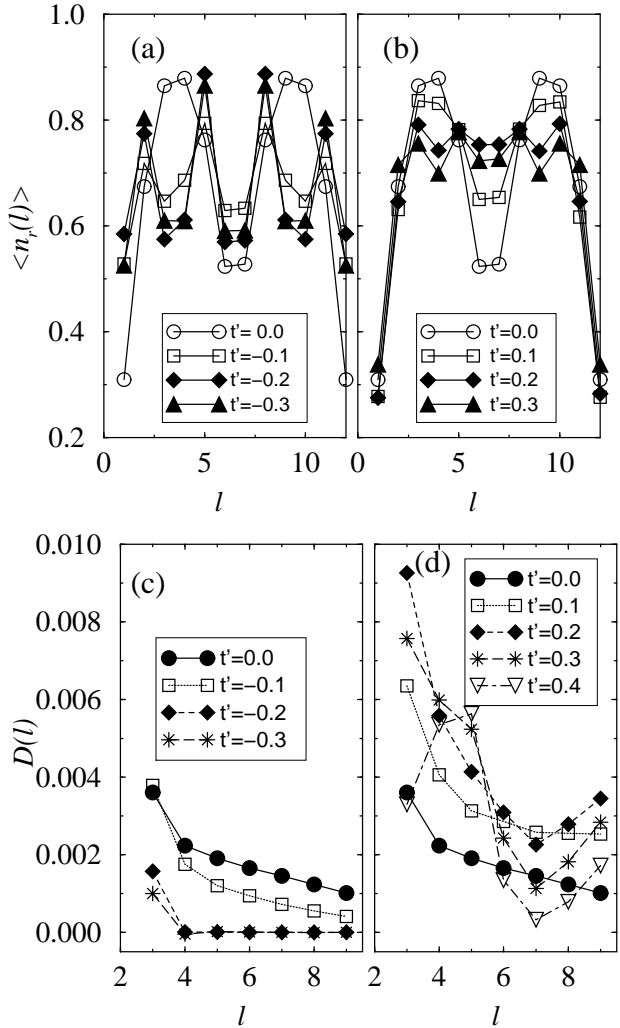}}
\caption{(a) Hole density per rung for a $12\times4$ system
with 8 holes, $J/t=0.35$ and open boundary conditions, with
$t' <= 0$. (b) Same as in (a), but
with $t' >= 0$. (c) and (d) $d$-wave pairing correlations for
the systems shown in (a) and (b), respectively.
}
\end{figure}

For this same $12\times 4$ lattice, we have studied the pair-field
correlation function
\begin{equation}
D(\ell) = \left\langle\Delta_{i+\ell} \Delta^+_i\right\rangle
\label{onethree}
\end{equation}
with $\Delta^+_i$ a pair creation operator which creates a singlet
$d_{x^2-y^2}$ pair centered on
the $i^{\rm\,th}$ site of the second leg.
Figure 1(b) shows a plot of $D(\ell)$ versus $\ell$ for the $4\times 12$
ladder for $J/t=0.35$ with 8 holes and various values of $t^\prime/t$. As
$t^\prime/t$ initially increases, the pairing correlations are enhanced but
as $t^\prime/t$ becomes greater than $\sim 0.3$, they are suppressed.  They are
suppressed for $t^\prime$ negative, with very strong suppression 
occuring for $t^\prime \le -0.2$.

Results for the charge density and spin structure of a $12\times 6$ lattice
with $J/t = 0.5$ and 8 holes are shown in Figure 2. Here we have taken
cylindrical boundary conditions, i.e. periodic in the $y$-direction, open
in the $x$-direction.
In this case, for $t^\prime/t=0$, the holes
form two transverse domains each containing 4 holes. The $\pi$-phase
shifted antiferromagnetic regions which are separated by these domains are
clearly visible in Figure 2 for $t'=0$. The DMRG calculation has selected a
particular spin order, breaking symmetry; 
as the number of states kept per block increases,
the magnitude of this spin order decreases, and the exact ground
state would have no net spin on any site. However, here the
spin order serves to illustrate the underlying spin correlations
in the exact ground state, which we expect to be a superposition
of the broken symmetry state rotated to all possible directions.

\begin{figure}[ht]
\epsfxsize=2.8 in\centerline{\epsffile{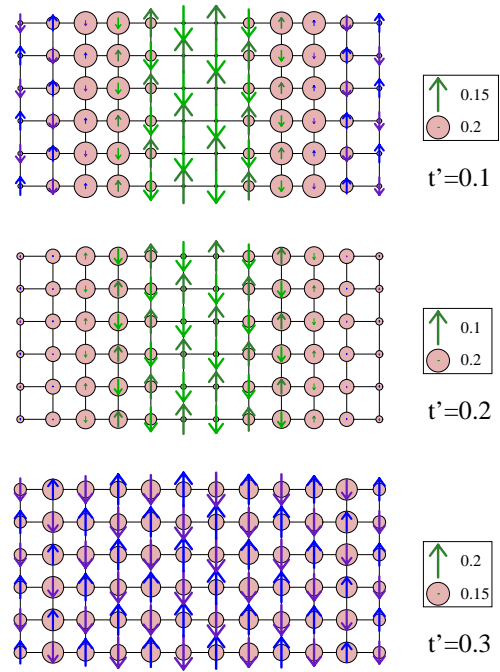}}
\caption{Hole and spin densities on $12\times 6$ systems
with cylindrical boundary conditions. The hole density is
proportional to the diameter of the circles, according to the
indicated scale, and similarly the length of the arrows gives
the expectaiton value of $S_z$.
}
\end{figure}

As $t^\prime/t$ increases, we again see a suppression of the
charge order and in addition the $\pi$-phase shifted antiferromagnetic
regions disappear. This is also true for $t^\prime$ negative.
For $t'=0.3$, we see that N\'eel spin order, without any
$\pi$ phase shifts, is now the broken symmetry state.
As previously noted, Lanczos \cite{TM94} and Monte Carlo
\cite{DM95}
calculations indicated
that a positive $t^\prime$ tended to stabilize the
commensurate $(\pi, \pi)$ antiferromagnetic correlations, which
is consistent with our results.

The rung density shown in Figure 3(a) provides a more
quantitative display of the suppression of the charge domains walls. 
In this case, a finite
magnitude of $t'$ seems to be necessary to substantially reduce
the charge density structure.  The domain walls in $L\times6$
ladders at $t'=0$ are
stable bound states of two hole pairs, and a finite change in
the parameters of the systems is needed to break them up. We
believe that $L\times6$ cylindrical systems have unusually stable domain
walls, and that more generally a smaller value of $|t'|$ would
destabilize the stripes.
Here, we see that the stripes are suppressed for $t'=0.2$,
and completely destabilized for $t'=0.3$.

Figures 3(c) and (d) show the pair-field 
correlations $D(\ell)$ versus $\ell$ for
various values of $t^\prime/t$ for the $12\times 6$ ladder. 
We see when the stripes are weakened by a positive $t'$, 
pairing correlations are strongly enhanced. The optimal $t'$
appears to be near $t'=0.2$.
Pairing is once again suppressed for negative $t'$, even when
the domain walls are destabilized.

From a weak coupling point of view, our results on the effect
of $t'$ on pairing are surprising.
In weak coupling, the effect of $t' < 0$ is to shift the van
Hove singularity in the density of states away from
half-filling, so that the singularity may occur near the Fermi level
in a doped system.  Thus, one might have expected to find
an enhancement in pairing for $t' < 0$.
However, in the $t-J$ model, we find a suppression of the
pairing.  In strong coupling, one can understand this effect.
Consider a pair
of holes, and imagine we fix one hole and let the other hole hop
around it. Consider the phase of the wavefunction of the second
hole on the four sites next to the first hole.  It appears that
$t' < 0$ will directly favor a +-+- $d$-wave phase pattern as
the second hole hops around the first, whereas $t' > 0$ would
favor the $s$-wave pattern ++++\cite{nazarenko}. 
However, the actual phase of a
pair is a relative phase between a system with $N$ holes and one
with $N+2$ holes.  If one considers a $2\times 2$ $t$-$J$
system, one finds that the 2-hole ground state has $s$-wave 
rotational symmetry, whereas the undoped state has $d$-wave
rotational symmetry\cite{moreo,trugman,holestructures}. 
The $d$-wave nature of the pairing comes from the difference in
these rotational symmetries.  Consequently, $t' < 0$,
by suppressing the 2-hole ++++ pattern, actually suppresses
$d$-wave pairing, while $t' > 0$ can enhance it.

\begin{figure}[ht]
\epsfxsize=3.3 in\centerline{\epsffile{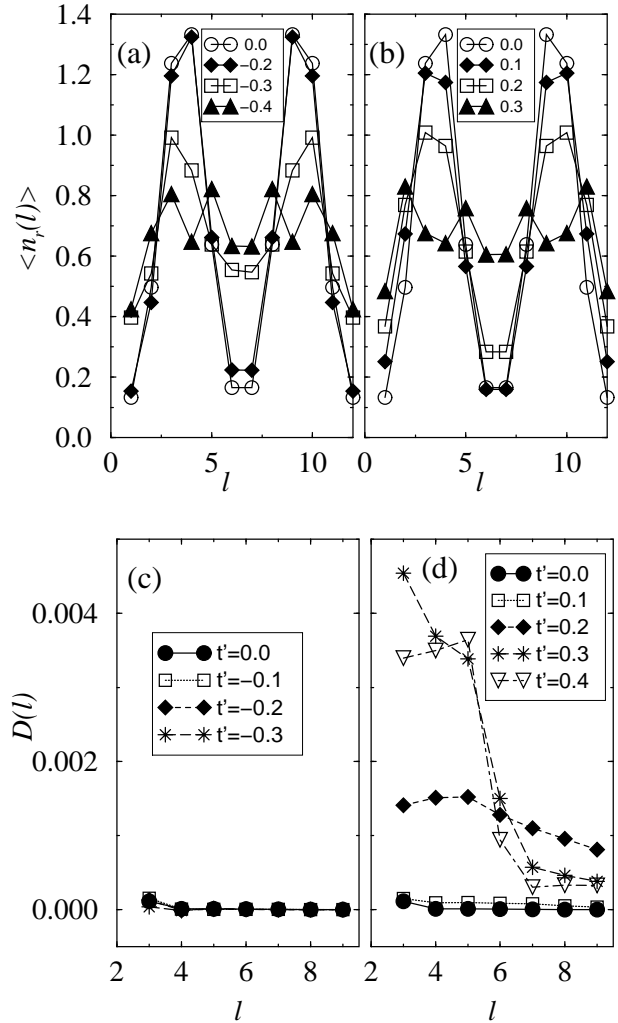}}
\caption{(a) Density of holes per rung for the $12\times 6$
ladder systems shown in
Fig. 2. (b) $d$-wave pairing correlations for
the same systems.
}
\end{figure}

Consider the $2 \times 2$ system\cite{holestructures}.   The
energy of the undoped system is independent of $t'$; we find
$E(0) = -3 J$. The energy of the one hole system depends only
weakly on $t'$; for $t'$ small, we find $E(1) = -J - 1/2 (J^2 +
12 t^2 + 4 J t' + 4 t'^2)^{1/2}$.  For $J = 0.35$, $t=1$, this
varies with $t'$ as $E \approx -2.09087 - 0.1005 t'$. The energy
of the two-hole system, in contrast, depends strongly on $t'$:
$E(2) = -J/2 - t' - (32 t^2 + (J + 2t')^2)^{1/2}$.  The pair
binding energy is defined as 
\begin{equation} 
E_b = 2 E(1) - E(2) - E(0).  
\end{equation} 
The dependence of the pair binding
energy on $t'$ is dominated by $E(2)$, and we find that $t' > 0$
strongly enhances the pair binding.

On larger systems, the detailed energetics are more complex, but
a similar effect occurs. In Fig. 4, we show the energy per hole
of several systems as a function of $t'$\cite{energetics}. 
The systems allow us to compare the stability of paired
states, striped states, and states with isolated holes.
The first system is
a single hole in an $8\times8$ open system, with a staggered
antiferromagnetic field of strength 0.1 on the edges to
approximate the magnetic coupling to
the rest of the system, which is assumed to be undoped.
The second system is similar, but has two holes. We plot the
energy difference between these systems and the same system
without holes, divided by the number of holes\cite{energetics}. 
The third system is a $16\times6$ system, with open boundary
conditions, and staggered fields of magnitude 0.1
with a $\pi$ phase shift applied on the first and last chain.
These boundary conditions favor the development of a stripe down
the center of the ladder. Then we subtract the energy of an
undoped $16\times6$ system, also with staggered fields, but
without the phase shift\cite{energetics}. We expect that finite
size effects are not neglible, and these could 
shift the striped phase
curve relative to the other two curves. However, we believe the general
trends are reliable. That is,
the striped system is lowest in energy near $t'=0$, but becomes
unstable as $t'$ becomes less than $-0.1$, or as $t'$ increases
above a value slightly greater than $0.0$. Thus, the striped
region is quite narrow as a function of $t'$.
This conclusion differs somewhat from that of Ref.
\cite{tohyama}, where it was found that stripes were enhanced
for $0<t'<0.2$, but were suppressed for larger values of $t'$.
For positive $t'$, the new stable state has pairs of holes,
as Tohyama, et. al.\cite{tohyama} also found for $t' \sim 0.5$.
For $t' < -0.1$, the near degeneracy between one and two holes
indicates that the holes are not bound into pairs: instead,
the stripes break up into quasiparticles. These
observations are consistent with enhanced pairing correlations for $t'>0$,
and suppressed pairing correlations for $t'<0$. Note that

\begin{figure}[ht]
\epsfxsize=2.8 in\centerline{\epsffile{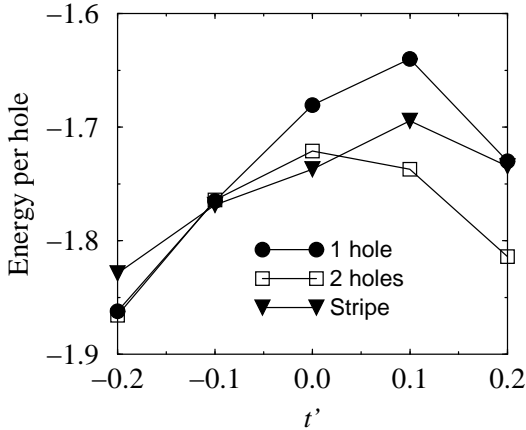}}
\caption{Energy per hole of various hole configurations, as
discussed in the text. 
}
\end{figure}

$N$-leg, $t-J$ ladders provide perhaps the simplest models which exhibit
many phenomenologically similar characteristics to those observed in the
cuprates. Here, for two different $t-t^\prime-J$ ladders, we find that a
diagonal, next-near-neighbor hopping suppresses the formation of static
stripes and that for $t^\prime > 0$
this can lead to an enhancement of the $d_{x^2-y^2}$
pairing correlations, while $t'<0$ we find suppression of
pairing. 


We thank M.P.A.~Fisher, S.A.~Kivelson, A.~Millis, S.~Sachdev,
and E.~Dagotto
for interesting
discussions. D.J.~Scalapino would like to acknowledge the Aspen Center for
Physics where the interplay of stripes and superconductivity in these
models was discussed. S.R.~White acknowledges support from the NSF under
grant \# DMR98-70930 and D.J.~Scalapino acknowledges support from the NSF
under grant \# DMR95-27304.

\end{document}